\documentclass{article}
\usepackage[T1]{fontenc} 
\usepackage[utf8]{inputenc} 
\usepackage{ismir,amsmath,cite,url}
\usepackage{graphicx}
\usepackage{color}
\usepackage[dvipsnames]{xcolor}
\usepackage{lineno}

\def\final{}

\iffalse
    \newcommand{\davide}[1]{{\color{cyan}[DAVIDE: #1]}}
    \newcommand{\giulio}[1]{{\color{red}[GIULIO: #1]}}
    \newcommand{\jordi}[1]{{\color{brown}[JORDI: #1]}}
    \newcommand{\dani}[1]{{\color{green}[DANI: #1]}}
    \newcommand{\santi}[1]{{\color{orange}[SANTI: #1]}}
    \newcommand{\joan}[1]{{\color{blue}[JOAN: #1]}}
\else
    \newcommand{\davide}[1]{}
    \newcommand{\giulio}[1]{}
    \newcommand{\jordi}[1]{}
    \newcommand{\dani}[1]{}
    \newcommand{\santi}[1]{}
    \newcommand{\joan}[1]{}
\fi

\usepackage{amssymb}
\usepackage{pifont}
\newcommand{\cmark}{\ding{51}}
\newcommand{\xmark}{\ding{55}}
\newcommand{\ma}[1]{\textbf{#1}}
\newcommand{\ve}[1]{\textbf{#1}}
\newcommand{\te}[1]{\bm{\mathcal{#1}}}

\newcommand{\tabsp}{@{\hskip 0.3in}}
\ifdefined\final
    \newcommand{\captionspacing}{\vspace{-0.2cm}}
\else
    \newcommand{\captionspacing}{}
\fi

\ifdefined\final
\else
\linenumbers
\fi

\title{Mono-to-stereo through parametric stereo generation}

\ifdefined\final
    \multauthor
    {Joan Serr\`a \hspace{0.5cm} Davide Scaini \hspace{0.5cm} Santiago Pascual \hspace{0.5cm} Daniel Arteaga}{ \bfseries{Jordi Pons \hspace{0.5cm} Jeroen Breebaart \hspace{0.5cm} Giulio Cengarle}\\
    \vspace{-0.3cm} ~\\ Dolby Laboratories \\
    {\tt\small firstname.lastname@dolby.com}}
\else
    \threeauthors
      {First Author} {Affiliation1 \\ {\tt author1@ismir.edu}}
      {Second Author} {\bf Retain these fake authors in\\\bf submission to preserve the formatting}
      {Third Author} {Affiliation3 \\ {\tt author3@ismir.edu}}
\fi

\ifdefined\final
    \def\authorname{J.~Serrà, D.~Scaini, S.~Pascual, D.~Arteaga, J.~Pons, J.~Breebaart, and G.~Cengarle}
\else
    \def\authorname{F. Author, S. Author, and T. Author}
\fi

\usepackage[bookmarks=false,pdfauthor={\authorname},pdfsubject={\papersubject},hidelinks]{hyperref}

\begin{document}

\maketitle

\begin{abstract}
Generating a stereophonic presentation from a monophonic audio signal is a challenging open task, especially if the goal is to obtain a realistic spatial imaging with a specific panning of sound elements. In this work, we propose to convert mono to stereo by means of predicting parametric stereo (PS) parameters using both nearest neighbor and deep network approaches. In combination with PS, we also propose to model the task with generative approaches, allowing to synthesize multiple and equally-plausible stereo renditions from the same mono signal. To achieve this, we consider both autoregressive and masked token modelling approaches. We provide evidence that the proposed PS-based models outperform a competitive classical decorrelation baseline and that, within a PS prediction framework, modern generative models outshine equivalent non-generative counterparts. Overall, our work positions both PS and generative modelling as strong and appealing methodologies for mono-to-stereo upmixing. A discussion of the limitations of these approaches is also provided.
\end{abstract}

\section{Introduction}
\label{sec:intro}

Single-channel monophonic (mono) signals are found in multiple situations, such as historical recordings or current ones made with a single microphone (e.g.,~field recordings, amateur band rehearsals, etc.). Even recordings made with two or more microphones that are not spaced enough or that do not have enough directivity may be better treated by downmixing to mono (e.g.,~mobile phone recordings). Furthermore, many processing algorithms, including modern deep neural network algorithms, cannot yet or are simply not designed to handle more than one channel. 
Unlike these scenarios, the most common listening experiences, either though loudspeakers or headphones,  involve two-channel stereophonic (stereo) signals. Hence the usefulness of mono to stereo upmixing.

Classical approaches to produce a pseudo-stereo effect from a mono signal are based on decorrelation. Initial approaches used time delays and complementary filters~\cite{Schroeder_1958}, although all-pass filters~\cite{Bauer_1963} are commonly used nowadays, together with multi-band processing to improve the effect~\cite{Orban_1970,Faller_2005,Fink_2015}. Instead of multi-band, estimation of foreground/background time-frequency tiles can also be performed~\cite{Uhle_2016}. Decorrelation approaches, however, only provide a mild stereo effect, with limited width, and cannot spatially separate individual elements in the mix. To overcome the latter, researchers have considered source separation approaches~\cite{Lagrange_2007,FitzGerald_2011,DelgadoCastro_2019}. The main idea is that, if individual elements or tracks are available, those can be panned to any location, producing a more realistic spatial image. Nevertheless, this approach presents several drawbacks: firstly, even the best-performing source separation algorithms produce artifacts~\cite{Pons_2021}, which can be highly audible in the stereo render; secondly, current separation algorithms are very restrictive in the number and types of elements they can separate~\cite{cano2018musical}, thus considerably limiting their application in real-world spatialization tasks; thirdly, after elements or tracks are separated, it remains to be seen how can they be automatically panned in a realistic 
\ifdefined\final
manner (cf.~\cite{Steinmetz_2021}), 
\else
manner, 
\fi
which is the reason why separation-based approaches usually involve user intervention in the panning stage~\cite{Lagrange_2007,FitzGerald_2011,DelgadoCastro_2019}. 

Music is a paradigmatic example where, apart from stereo capture, artists and engineers massively exploit the stereo image to serve a creative artistic intent. Instrument panning is a fundamental part of music mixing, and achieving the right balance requires musical sensibility as well as technical knowledge~\cite{Gibson_2005}. However, apart from some style conventions, the stereo image of a music mix is a highly subjective construct: given a set of input tracks, there are many plausible stereo renditions from which selecting the final mix is practically only a matter of artistic choice. Hence, we posit that this is a perfect ground for modern deep generative models~\cite{Tomczak_2022}. However, to our surprise, we only found one work using deep neural networks for mono-to-stereo~\cite{Chun_2015}, with very limited generative capabilities.

In this work, we propose the use of machine learning techniques and parametric stereo (PS) decoding~\cite{Purnhagen_2004,Breebaart2005} for converting mono to stereo. PS is a coding technique that allows to transmit a stereo signal through a mono signal plus side information that, with enough bit rate, can be used to recover an almost transparent version of the original stereo content. 
By leveraging machine learning techniques, we generate (or invent) plausible versions of PS parameters in situations where side information is not available. These parameters can then be used to decode an existing mono signal into a plausible stereo one. 
We propose two variants of PS generation: one based on a classical nearest neighbor approach~\cite{Hastie_2009} and another one based on deep generative modeling. For the latter, we consider both common autoregressive modeling~\cite{Radford_2018} and more recent masked token modeling~\cite{Chang_2022}, and show that there can be noticeable differences between the two. We use subjective testing to compare the proposed approaches and show that PS generation can produce results that are more appealing than considered competitive baselines. We also introduce two objective evaluation metrics and discuss the limitations of both PS and generative approaches for mono-to-stereo.

\section{Parametric stereo}
\label{sec:ps}

PS exploits the perceptual cues that are more relevant to our spatial perception of sound, namely the fact that directional sources produce interaural level and phase (or time delay) differences, and the fact that diffuse sound fields manifest as decorrelated signals at the two ears. These cues effectively describe how a mono signal is mapped to the left and right stereo channels, and can be measured using three quantities or parameters~\cite{Purnhagen_2004,Breebaart2005}: interchannel intensity differences (IID), interchannel time differences (or, equivalently, phase differences), and interchannel coherence or correlation (IC). 
PS parameters are computed in frequency bands, to reflect the frequency-dependent nature of the spatial properties of stereo content, and also on a frame-by-frame basis, to reflect the time-varying nature of frequency cues and spatial images. 
An important observation is that PS is capable of capturing spatial attributes that are perceptually relevant and re-instate those without changing signal levels, tonality, or other artifacts that may arise from methods that operate on audio signals directly. 
In this work, for compactness and ease of implementation, 
we choose to use the two-parameter approach by Breebaart~et~al.~\cite{Breebaart2005}, which models IID and IC without interchannel phase differences, accepting that this two-parameter approach is not providing the best possible quality of PS coding. We now overview this PS coding strategy and introduce the main notation of the article.

\subsection{Encoding}
\label{sec:ps_encode}

Given two complex-valued spectrograms expressed as complex matrices $\ma{X}$ and $\ma{Y}$, where rows represent frequency bins and columns represent frames, we define the band-based cross-spectrogram function
\begin{equation*}
    \rho(\ma{X},\ma{Y}) = \ma{B}~(\ma{X}\odot\ma{Y}^*) ,
\end{equation*}
where $\odot$ denotes elementwise multiplication, $^*$ denotes elementwise complex conjugate, and $\ma{B}$ is a matrix with ones and zeros that is used to sum frequency bins according to a certain frequency band grouping (using matrix multiplication). In this work, we use the same spectrogram settings and banding as in~\cite{Breebaart2005}: frames of 4,096~samples for 44.1\,kHz signals, 75\% overlap, a Hann window, and 34~bands which are approximately distributed following equivalent rectangular bandwidths. 

Given the two complex spectrograms $\ma{L}$ and $\ma{R}$ corresponding to the left and right channels of a stereo signal, we can compute the IID using
\begin{equation*}
    \ma{P}^{\text{IID}} = 10\log_{10} \left( \rho(\ma{L},\ma{L}) \oslash \rho(\ma{R},\ma{R}) \right),
\end{equation*}
where $\oslash$ denotes elementwise division. The IC is similarly derived from the cross-spectrogram following
\begin{equation*}
    \ma{P}^{\text{IC}} = \text{Re}\left\{ \rho(\ma{L},\ma{R}) \right\} \oslash \sqrt{ \rho(\ma{L},\ma{L})\odot\rho(\ma{R},\ma{R}) },
\end{equation*}
where $\text{Re}\{\}$ extracts the real part of each complex value and the square root is applied elementwise. Notice that the use of the real part instead of the absolute value allows to retain information on the relative phase of the two signals that would otherwise be lost.
We finally quantize $\ma{P}^{\text{IID}}$ and $\ma{P}^{\text{IC}}$ by discretizing each matrix element. To do so, we use the same non-uniform quantization steps as in~\cite{Breebaart2005}: 31~steps for IID and 8 for IC. We denote the quantized versions as $\ma{Q}^{\text{IID}}$ and $\ma{Q}^{\text{IC}}$.

To facilitate subsequent operation, and to prevent potential prediction mismatches between IID and IC, we join both parameters and treat them as one. For $\ma{P}^{\text{IID}}$ and $\ma{P}^{\text{IC}}$, we concatenate them in the frequency axis and form a single matrix $\ma{P}$. For $\ma{Q}^{\text{IID}}_{i,j}$ and $\ma{Q}^{\text{IC}}_{i,j}$, we fuse them elementwise into individual integers using the amount of IC quantization steps. This way, $\ma{Q}_{i,j}=8\cdot\ma{Q}^{\text{IID}}_{i,j}+\ma{Q}^{\text{IC}}_{i,j}$ (note that we can recover back $\ma{Q}^{\text{IID}}_{i,j}$ and $\ma{Q}^{\text{IC}}_{i,j}$ using the division and modulo operators). 

\subsection{Decoding}

To decode the above PS encoding, we perform a mixing between the available mono signal and a decorrelated version of it. We decorrelate a mono signal $\ma{S}$ by applying a cascade of 4~infinite impulse response all-pass filters and obtain $\ma{S}^{\text{D}}$ (this all-pass filter is an enhanced version of the basic one proposed in~\cite{Breebaart2005} thanks to transient detection and preservation, which avoids time smearing). After that, we can decode the estimated left and right channels $\hat{\ma{L}}$ and $\hat{\ma{R}}$ by carefully mixing $\ma{S}$ and $\ma{S}^{\text{D}}$. We can do so with
\begin{equation*}
    \begin{array}{c}
        \hat{\ma{L}} = \ma{M}^{\text{a}}\odot\ma{S} + \ma{M}^{\text{b}}\odot\ma{S}^{\text{D}} , \\
        \hat{\ma{R}} = \ma{M}^{\text{c}}\odot\ma{S} + \ma{M}^{\text{d}}\odot\ma{S}^{\text{D}} , \\
    \end{array}
\end{equation*}
using mixing matrices $\ma{M}$, which are computed from the coded PS parameters $\ma{P}^{\text{IID}}$ and $\ma{P}^{\text{IC}}$. The exact calculation of mixing matrices $\ma{M}$ is straightforward to obtain by adapting to matrix notation the formulation in~\cite{Breebaart2005}, to which we refer for further detail and explanation.

\section{Parametric stereo generation}
\label{sec:psgen}

We now explain the proposed approaches for PS generation. All of them share the above encoding-decoding formulation, either using the quantized or unquantized versions. During training, stereo signals are used to compute input downmixes $\ma{S}=(\ma{L}+\ma{R})/2$ and target PS parameters $\ma{P}$ or $\ma{Q}$ (hence the proposed approaches aim at producing $\hat{\ma{P}}$ or $\hat{\ma{Q}}$). Note that, in the case of the generative models we consider, one has to additionally input contextual PS parameters in a teacher-forcing schema~\cite{Williams_1989}. We also want to note that, since they are quite common practice, it is not in the scope of the current work to provide a detailed explanation of existing generative models (instead, we refer the interested reader to the cited references). In all proposed approaches, we tune model hyperparameters by qualitative manual inspection in a preliminary analysis stage. PS specifications are predefined and correspond to the ones mentioned in Sec.~\ref{sec:ps}. Neural network approaches use Pytorch's~\cite{Paszke_2019} defaults and are trained with Adam for 700~epochs using a batch size of 128 and a learning rate of $10^{-4}$, with warmup cosine scheduling. 

\subsection{Nearest neighbor}
\label{sec:nneigh}

The first approach proposes to impose the PS parameters of existing, similar stereo fragments to individual mono frames using a nearest neighbor (NN) algorithm~\cite{Hastie_2009}. We call the approach PS-NN. The idea is to retrieve frame-based PS parameters using mono frame sequences, and to use the sequence of those retrieved parameters to decode the mono input. 
At training time, we randomly select a song, randomly extract an $N=20$~frame spectrogram $\ma{S}$ and its corresponding parameters $\ma{P}$, and compute a key-value vector pair (we here use the magnitude spectrogram). The key vector is formed by framewise averaging the energy in each band,
\begin{equation}
    \ve{k}=\frac{1}{N}\sum_{j=1}^{N}\ma{B}\,\ma{S}_{:,j} ,
    \label{eq:nn_keys}
\end{equation} 
and the value vector corresponds to the PS parameters of the last frame, $\ve{v}=\ma{P}_{:,N}$, which allows for a fully-causal schema. We repeat the process half a million times and store all pairs in a nearest neighbor structure. At test time, for every frame of the input mono signal, we compute an average as in Eq.~\ref{eq:nn_keys}, query the nearest neighbor structure, retrieve the $\hat{\ve{v}}$ vector of the closest neighbor (using Euclidean distance), and assign it as the predicted PS parameter for that frame. This way, we obtain a sequence of estimated PS parameters $\hat{\ma{P}}$.

In preliminary analysis, we observed that PS-NN produced a high-rate `wobbling' effect between left and right (that is, panning was rapidly switching from one channel to the other) and presented some temporal inconsistencies (that is, sources were unrealistically moving with time, even within one- or two-second windows). To counteract these effects, we implemented a two step post-processing based on (i)~switching the sign of $\hat{\ma{P}}{}^{\text{IID}}_{:,j}$ if the Euclidean distance to $\hat{\ma{P}}{}^{\text{IID}}_{:,j-1}$ was smaller, and (ii)~applying an exponential smoothing on the columns of $\hat{\ma{P}}$ with a factor of 0.95. This post-processing substantially reduced the aforementioned undesirable effects.

\subsection{Autoregressive}
\label{sec:ar}

The second approach proposes to model PS parameters with a deep generative approach based on an autoregressive (AR) transformer~\cite{Radford_2018}. We call the approach PS-AR. Our architecture is composed by 7~transformer encoder blocks of 512~channels, with 16~heads and a multilayer perceptron (MLPs) expansion factor of 3. We use sinusoidal positional encoding at the input, and add a two-layer MLP with an expansion factor of 2 at the output to project to the final number of classes (which is 31$\times$8~tokens times 34~bands per frame, see Sec.~\ref{sec:ps_encode}). The input is formed by a projection of the mono spectrogram $\ma{S}$ and the teacher-forcing information $\ma{Q}$ into a 512-channel activation $\ma{H}$,
\begin{equation}
    \ma{H} = \phi(\ma{S}) + \sum_{i=1}^{B} \xi_i(\ma{Q}_{i,:}) ,
    \label{eq:dnn_input}
\end{equation}
where $\phi$ is a two-layer MLP with an expansion factor of~2, $B=34$ is the number of bands, and $\xi_i$ is a learnable per-band token embedding (which includes the mask token, see below). We train the model with weighted categorical cross-entropy, using the weight
\begin{equation}
    w = 1 + \lambda\sigma\left( \left[\ma{P}^{\text{IID}}\right]_{\pm\epsilon} \right) + \sigma(\ma{P}^{\text{IC}}) ,
    \label{eq:ce_weight}
\end{equation}
calculated independently for every element in the batch. In Eq.~\ref{eq:ce_weight}, $\sigma(\ma{X})$ corresponds to the elementwise standard deviation of $\ma{X}$, $\lambda=0.15$ compensates for different magnitudes, $[~]_{\pm\epsilon}$ corresponds to the clipping operation, and $\epsilon=20$ is a threshold to take into account the little perceptual relevance of IIDs larger than 20\,dB~\cite{Bartlett_1991}. In preliminary analysis, we observed that using $w$ qualitatively improved results, as it shall promote focus on wider stereo images and more difficult cases. 

PS-AR follows a PixelSNAIL recursive approach~\cite{Chen_2018}, starting with the prediction of lower frequency bands, then higher frequency bands, and moving into the next frame once all bands are predicted. To efficiently exploit the past context, all input sequences have full-sequence teacher-forcing except for the upper frequency bands of the last frame, which are masked consecutively and uniformly at random during training~\cite{Chen_2018}. At test time, we sample recursively, following the same masking strategy and using a temperature hyperparameter $\tau=0.9$. In addition, we employ classifier-free guidance~\cite{Ho_Salimans_2021} with a hyperparameter $\gamma=0.25$. For that, we use the approach in~\cite{Chang_2023}, which modifies the conditional logits $\ma{U}^{\text{cond}}$ with unconditional ones $\ma{U}^{\text{uncond}}$ such that
\begin{equation}
    \ma{U} = (1+\gamma) \ma{U}^{\text{cond}} - \gamma \ma{U}^{\text{uncond}} .
    \label{eq:guidance}
\end{equation}
To have both a conditional and an unconditional model within the same architecture, following common practice, we randomly replace $\phi(\ma{S})$ in Eq.~\ref{eq:dnn_input} by a learnable dropout token 10\% of the time.

\subsection{Masked token modeling}
\label{sec:mtm}

The third approach proposes to model PS parameters with a deep generative approach based on masked token modeling (MTM)~\cite{Chang_2022}. We call the approach PS-MTM. The architecture, loss, inputs, and outputs of the model are the same as in PS-AR, including the cross-entropy weights (Eq.~\ref{eq:ce_weight}) and classifier-free guidance (Eq.~\ref{eq:guidance}). The only difference is the masking approach and the sampling procedure, which implies different hyperparameters for the testing stage (we use $\tau=4.5$ and $\gamma=0.75$, but now the temperature $\tau$ has a different meaning as explained below).

MTM generates patch representations $\ma{Q}$ with quantized elements $\ma{Q}_{i,j}$ which are dubbed as tokens (in our case the matrix $\ma{Q}$ has dimensions $B\times N$, with $N$ being the number of considered audio frames; the maximum number of tokens in $\ma{Q}_{i,j}$ is 31$\times$8, as defined in Sec.~\ref{sec:ps_encode}). During training, the teacher-forcing input $\ma{Q}$ is masked uniformly at random, and only the ground truth elements corresponding to the masked positions are used to compute the cross-entropy loss at the output. The number of elements to mask is also selected at random following a cosine schedule~\cite{Chang_2022} (this specifically includes the case where all patch elements are masked). During sampling, patch representations are formed with 50\% overlap, using no masking for the first half of the patch, similar to~\cite{Chang_2023}.

MTM sampling is an iterative process that achieves orders of magnitude speedups compared to autoregressive modeling (in our case PS-MTM uses 20~steps for a 3\,s hop, while PS-AR requires $B=34$~steps for just a single audio frame of a few milliseconds). MTM iteratively samples a masked patch, performs predictions with classifier-free guidance~\cite{Chang_2023}, chooses the predictions with the highest logit score for the next iteration (they will become unmasked and fixed), and reduces the percent of masked tokens following the same scheduling as in training until no masked elements remain~\cite{Chang_2022}. Differently from training, the masking used in sampling is not random, but based on logit scores (lowest ones become masked), and noise is added to logit scores to promote diversity~\cite{Chang_2022,Chang_2023}. In our case, we employ Gaussian noise with zero mean and a standard deviation $\tau$, which becomes our redefined temperature parameter. \joan{Santi, did I forget anything relevant for AR or MTM?}

\section{Evaluation}
\label{sec:eval}

To train and evaluate all approaches we use a collection of professionally-recorded stereo music tracks at 44.1\,kHz. We consider 419,954~tracks for training and 10\,k for evaluation, and randomly extract a 10\,s chunk from each track. During training, we sample 6\,s patches from those and perform data augmentation using a random gain and also randomly switching left and right channels.

\subsection{Baselines: regression and decorrelation}

In addition to the original stereo and its mono downmix, we consider two additional baselines to compare with the previous approaches. 
The first baseline corresponds to an ablation of the deep generative approaches, and tries to answer the question of whether a generative component is needed or convenient for the task. Thus, the baseline consists of a neural network with the exact same configuration as PS-AR or PS-MTM, but substituting the generative part by standard regression with mean-squared error~\cite{Hastie_2009}. We term this baseline PS-Reg, and note that it could be considered an enhanced modern version of the approach of Chun~et~al.~\cite{Chun_2015}, using PS.

It is interesting to mention that, in preliminary analysis, we observed that PS-Reg accurately estimated IC values, but consistently failed to predict IIDs. The predicted IIDs had minimal deviation from zero, which can be attributed to the probability distribution function of IID values being centered around zero with equally plausible deviations to the right and to the left. 
This was an early indication that the one-to-many mapping of IID prediction cannot be correctly handled by regression methods, and that the task would be better served by a generative approach. 

The second baseline we consider corresponds to a variant of classical decorrelation approaches. Here, the decorrelation is implemented by means of an all-pass filter network enhanced by (i)~detection and preservation of transients, and (ii)~a frequency-dependent mix between original and decorrelated signals to achieve a frequency-dependent IC. We term this baseline Decorr, and we note that it could be considered an improved modern version of the approaches~\cite{Schroeder_1958,Bauer_1963,Orban_1970,Faller_2005,Fink_2015,Uhle_2016}. 

\begin{figure*}[!t]
 \centerline{
 \includegraphics[width=1\linewidth]{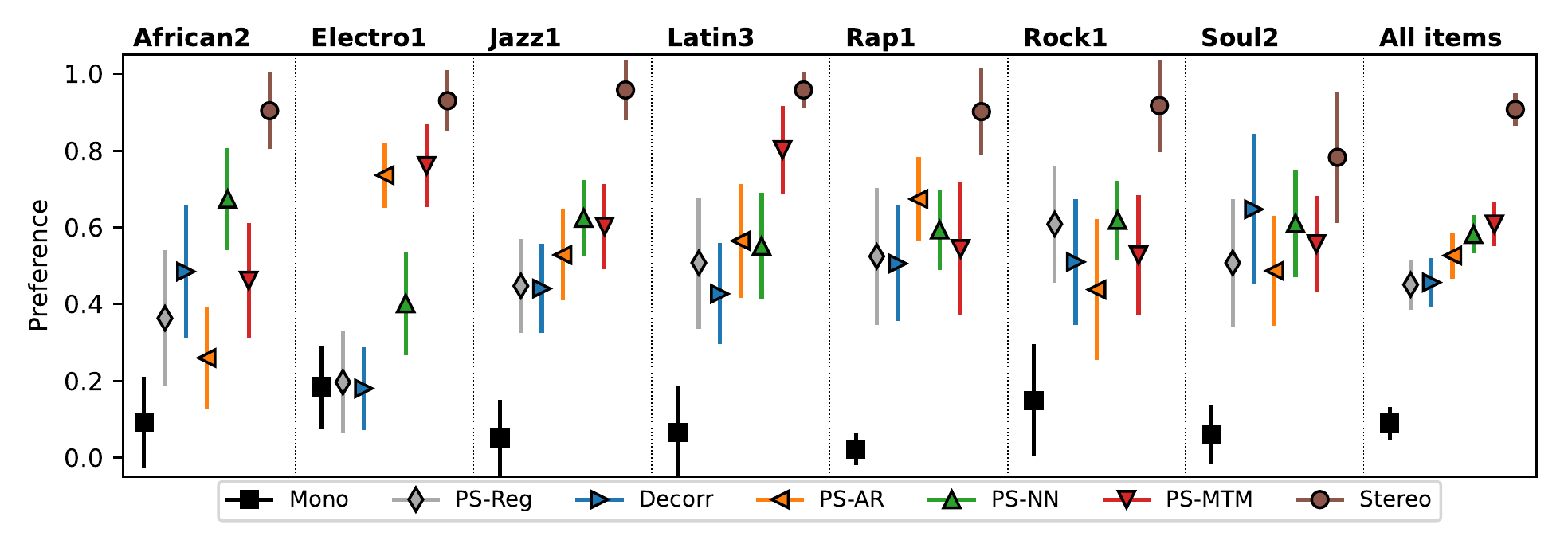}
 }
 \captionspacing
 \caption{Preference results for the items included in the subjective test (Sec.~\ref{sec:eval_subj}). Markers indicate average values and vertical bars indicate the 95\% confidence interval associated to them.}
 \label{fig:result}
\end{figure*}

\subsection{Objective measures}

To the best of our knowledge, there are no objective measurements for plausible stereo renderings nor suitable PS prediction scores. Due to the highly creative/subjective nature of the task, common error measurements may not be appropriate. Therefore, as a way of measuring progress, we propose to use a couple of metrics inspired from the literature on generative modeling (cf.~\cite{Tomczak_2022}). The first metric we consider is the minimum error on a large sample basis, $E_{\min}$. Given a large sample of generated PS parameters ($K=128$ for a single audio excerpt), $E_{\min}$ chooses the minimum error with respect to the ground truth:
\begin{equation*}
    E_{\min} = \min_k \left[ \sum_{i,j} \delta\left( \ma{P}_{i,j},\hat{\ma{P}}_{i,j}^{(k)} \right) \right] ,
\end{equation*}
where $\delta$ is a suitable error function. The idea is that if we allow the model to generate many samples for every input, in the limit of very large $K$ one of them should come close to the ground truth. For PS parameters, we use absolute errors, weight the IID to compensate magnitudes with IC, and take into account some perceptual relevance for IID as in Sec.~\ref{sec:ar} and Eq.~\ref{eq:ce_weight}:
\begin{equation*}
    \delta(x,y) = \begin{cases}
        \lambda\left| [x]_{\pm\epsilon} - [y]_{\pm\epsilon} \right| & \text{for IID}, \\
        \left| x-y  \right| & \text{for IC}. \\
    \end{cases}
\end{equation*}

The second metric we consider is the Fr\'echet distance on the PS parameter space, $D_{\text{F}}$. Given a pool of PS parameters $\te{P}$ and a $K$ times larger pool of generated parameters $\hat{\te{P}}$, assuming Gaussian distributions, $D_{\text{F}}$ is computed as
\begin{multline*}
    D_{\text{F}} = \left| \mu(\te{P}) - \mu(\hat{\te{P}}) \right|^2 + \\ + \text{Tr}\left\{ \sigma(\te{P}) + \sigma(\hat{\te{P}}) - 2\sqrt{\sigma(\te{P})\sigma(\hat{\te{P}})}  \right\},
\end{multline*}
where $\text{Tr}\{\}$ denotes the matrix trace and $\mu$ and $\sigma$ correspond to the mean vector and the covariance matrix over frames, respectively. The Fr\'echet distance has become a standard measure in generative modeling where, instead of the PS parameters used here, activations of pre-trained classification networks are used. We will see that it is also able to provide some informative cues in our task (Sec.~\ref{sec:result}).

\subsection{Subjective evaluation}
\label{sec:eval_subj}

Given the creative/subjective nature of the task, the best way to measure performance is through subjective testing. In this study, we ran a preference test with 24~listeners on 7~song excerpts of 10\,s from the test set. To select those excerpts, we ranked the test excerpts based on $w$ (Eq.~\ref{eq:ce_weight}) and randomly selected them from the top quartile. When doing so, 
we manually verified that the selected excerpts covered distinct musical genres and ensured that a PS-decoded version did not exhibit significant coding degradation 
(this way, we prime the listener to focus on the stereo image instead of potential artifacts introduced by our implementation of PS, Sec.~\ref{sec:ps}). 

The test consisted in providing a rating between 0 and 100 to 7~approaches: the three proposed ones, the two baselines, the mono downmix, and the original stereo signal (professional mix, non-coded). Mono and stereo signals provide us with intuitive bounds for the analysis of preference, 
and also serve us to discard non-expert listeners. Indeed, we found that the task is quite hard for non-experts, who provided many inconsistent ratings when asked to evaluate an appropriate balance between the width and the clarity of the mix. We used the most obvious of those inconsistencies to discard listeners from the test, namely the fact that they rated mono (input) over stereo (professional mix) in one or more occasions. Half of the users (12) did not incur into such inconsistency and were considered reliable enough to derive conclusions from their ratings. To compensate for differences in subjective scales, we normalized excerpt preference tuples between 0 and 1 (that is, we normalized the ratings for the 7~approaches independently per audio excerpt and listener). To measure statistical significance, we used pairwise Wilcoxon signed-rank tests and applied the Holm-Bonferroni adjustment for multiple testing with $p=0.05$. The Wilcoxon signed-rank test is appropriate for our case as it is non-parametric and designed for matched samples.

\section{Results}
\label{sec:result}

\joan{Are there interesting results that I have forgotten to mention? Discussion of limitations is in the next section} In Fig.~\ref{fig:result} we depict the average listener preference for each item and approach. Initially, we see that the pattern differs depending on the test item. For some items, the proposed approaches are preferred over the baselines (e.g.,~Electro1, Jazz1, and Latin3) while, for some other items, differences between approaches are less clear (e.g.,~Rock1 and Soul2). All approaches seem to be preferred above the mono signal, except for baseline approaches with Electro1. Noticeably, in some situations, preference for some of the proposed approaches even overlaps with the original stereo (e.g.,~Electro1, Latin3, and Soul2). The case of Soul2 shows an example where considered approaches are almost as preferred as the original stereo, whereas the case of Jazz1 shows an example where considered approaches are still far from the professional mix. 

Despite the different preferences on individual excerpts, upon further inspection we see that a clear pattern emerges when considering all items: proposed approaches rank better than mono and the considered baselines (Fig.~\ref{fig:result}, right). In \tabref{tab:results} we confirm that, on average, PS-AR is preferred over the baseline approaches and that, in turn, PS-NN and PS-MTM are preferred over PS-AR. In \tabref{tab:statsig}, we report statistically significant differences beetween PS-NN/PS-MTM and the baseline approaches, but not between PS-AR and the baseline approaches (and neither between PS-AR and PS-NN/PS-MTM nor between PS-NN and PS-MTM). Overall, the results show that a generative approach to PS prediction can become a compelling system for mono-to-stereo. The performance of PS-NN is a nice surprise that was not predicted by the objective metrics, which otherwise seem to correlate with listener preference (\tabref{tab:results}; perhaps PS-NN does not follow the trend because it is not a generative approach). 

\begin{table}[t]
 \begin{center}
 \begin{tabular}{l\tabsp cc\tabsp c}
  \hline\hline
  Approach & $E_{\min}$ $\downarrow$ & $D_{\text{F}}$ $\downarrow$ & Preference $\uparrow$ \\
  \hline\hline
  Mono      & 0.104 & 20.89 & 0.090 $\pm$ 0.042 \\
  PS-Reg   & 0.069 & 8.11 & 0.451 $\pm$ 0.066 \\
  Decorr   & 0.093 & 8.32 & 0.457 $\pm$ 0.064 \\
  \hline
  PS-AR    & 0.074 & 0.62 & 0.527 $\pm$ 0.060 \\
  PS-NN   & 0.089 & 3.08 & 0.582 $\pm$ 0.057 \\
  PS-MTM  & 0.068 & 0.59 & 0.608 $\pm$ 0.050 \\
  \hline
  Stereo    & 0.000 & 0.03 & 0.908 $\pm$ 0.042 \\
  \hline\hline
 \end{tabular}
\end{center}
 \captionspacing
 \caption{Results for the objective ($E_{\min}$, $D_{\text{F}}$) and subjective (Preference $\pm$ 95\% confidence interval) evaluations.}
 \label{tab:results}
\end{table}

\begin{table}[t]
 \begin{center}
 \setlength{\tabcolsep}{3pt}
 \resizebox{\linewidth}{!}{
 \begin{tabular}{l|cccccc}
  \hline\hline
      & PS-Reg & Decorr & PS-AR & PS-NN & PS-MTM & Stereo \\
  \hline
  Mono & \cmark & \cmark & \cmark & \cmark & \cmark & \cmark \\
  PS-Reg &  & \xmark & \xmark & \cmark & \cmark & \cmark \\
  Decorr &  &  & \xmark & \cmark & \cmark & \cmark \\
  PS-AR &  & & & \xmark & \xmark & \cmark \\
  PS-NN &  & & & & \xmark & \cmark \\
  PS-MTM &  & & & & & \cmark \\
  \hline\hline
 \end{tabular}
 }
\end{center}
 \captionspacing
 \caption{Pairwise statistical significance for the case of all test items 
 (12~subjects times 7~excerpts, see Sec.~\ref{sec:eval_subj}).
 The obtained $p$-value threshold is 0.0053. 
 }
 \label{tab:statsig}
\end{table}

Besides quality, another aspect worth considering is speed. In \tabref{tab:parinfer} we observe that PS-AR, as anticipated, is orders of magnitude slower than the other approaches, to the point of making it impractical for real-world operation. Decorr, PS-Reg, and PS-NN are faster than real-time on CPU and PS-MTM is not. However, one should note that with PS-MTM we can easily trade off sampling iterations at the expense of some quality reduction (see~\cite{Chang_2022,Chang_2023}). PS-NN may dramatically improve speed if we consider the use of fast nearest neighbor search algorithms or even hash tables, which make this approach very interesting for real-world deployment (note we deliberately made PS-NN comparable in size to the other approaches, see \tabref{tab:parinfer}).

\begin{table}[t]
 \begin{center}
 \begin{tabular}{l\tabsp c\tabsp cc}
  \hline\hline
  Approach & Learnable & \multicolumn{2}{c}{RTF $\downarrow$} \\
   & parameters & CPU & GPU \\
  \hline\hline
  Decorr  & 0 & 0.25  & n/a  \\
  PS-Reg   & 30.1\,M & 0.32  & 0.21  \\
  PS-NN   & 34.0\,M$^\dagger$ & 0.82 & n/a  \\
  PS-MTM  & 34.5\,M & 5.81  & 0.33  \\
  PS-AR    & 34.5\,M & 255.87 & 8.38  \\
  \hline\hline
 \end{tabular}
\end{center}
 \captionspacing
 \caption{Number of learnable parameters and average real-time factor (RTF). Superscript $^\dagger$ indicates an estimation of 0.5\,M key-value pairs with $B=34$ bands (Sec.~\ref{sec:nneigh}). RTFs are measured on a Xeon(R) 
 2.20\,GHz CPU and on a GeForce GTX 1080-Ti GPU. }
 \label{tab:parinfer}
\end{table}

\section{Discussion}
\label{sec:discuss}

Despite the good results obtained above, the subjective test reveals that, for some of the considered excerpts, there is still a gap between professional stereo mixes and the proposed approaches. We hypothesize that this gap is due to (i)~limitations of the considered PS encoding, and (ii)~the difficulty of the task itself. 
Regarding (i), we suspect that part of the low subjective scores of PS-based approaches is due to the audio distortions and tonal artifacts introduced by the PS decoding.
Thus, we hypothesize that using a commercial implementation of PS coding (or perhaps even learning end-to-end the coding operation) could yield better results. 
Besides, we think that the fact that PS is defined in a banded domain poses a challenge to PS generation approaches, namely that individual bands are panned but approaches do not have an explicit notion of instrument or `entity'. Indeed, we sometimes observe individual entities being panned into two different positions simultaneously (e.g.,~for the same instrument, we may get some frequencies panned to the left and some to the right, which is an uncommon stylistic decision). 
A potential solution to this problem could be to add better (or more) inputs to the models, together with more capacity, with the hope that they achieve a better understanding of what is a source before panning it. Along this line, it would be perhaps interesting to include some techniques used in the context of source separation with neural network models~\cite{cano2018musical}. 
Regarding (ii), another issue we sometimes observe is with the temporal consistency of panning decisions, with an instrument appearing predominantly in one channel but then moving (without much artistic criterion) to the other channel after 10 or 20\,s. Handling temporal consistency is a transversal problem across all generative models, typically handled by brute force (that is, more receptive field and/or larger models) or by some form of hierarchical or recurrent processing. Nonetheless, it is still an open issue, especially in the case of really long sequences like audio and music. 

In addition to the limitations inherent to the technology, there are also some shortcomings in the test methodology. The subjective tests were conducted using headphones, whereas stereo images are typically created and mixed in a studio using professional loudspeaker monitoring. This implies that when critically evaluating the proposed approaches on a professional setup, additional subtleties might be discernible.
Another methodological challenge was that often users had difficulty in evaluating multiple test excerpts according to the stated evaluation criteria. A potentially contributing factor to it was the absence of a standardized test methodology for multiple preference testing without a reference.

\section{Conclusion}

In this work we study methods to convert from mono to stereo. 
Our proposal entails (i)~the use of PS for mono to stereo upmixing and (ii)~the synthesis of PS parameters with three machine learning methods. We also introduce (iii)~the use of modern generative approaches to the task and propose two variants of them. 
We additionally (iv)~overview and adapt an existing PS methodology and (v)~propose two tentative objective metrics to evaluate stereo renderings. The three proposed approaches outperform the classical and the deep neural network baselines we consider, and two of such approaches stand out with a statistically significant difference in the subjective test. 

\ifdefined\final
\section{Acknowledgments}

We thank all the participants of the listening test for their input and Gautam Bhattacharya and Samuel Narv\'aez for preliminary discussions on the topic. 
\fi

\bibliography{mybib}

\end{document}